\documentstyle[prl,aps,multicol]{revtex}

\newcommand{\half}{{\textstyle{1\over2}}}

\newcommand{\beq}{\begin{equation}}
\newcommand{\eeq}{\end{equation}}
\newcommand{\bi}{\begin{itemize}}
\newcommand{\ei}{\end{itemize}}
\newcommand{\beqar}{\begin{eqnarray}}
\newcommand{\eeqar}{\end{eqnarray}}

\newcommand{\al}{\alpha}
\newcommand{\bb}{\beta}
\newcommand{\g}{\gamma}
\newcommand{\del}{\delta}
\newcommand{\e}{\epsilon}
\newcommand{\mn}{{\mu\nu}}
\newcommand{\pa}{\partial}
\newcommand{\npa}{\rlap{/}\partial}
\newcommand{\nA}{\rlap{/}\!A}

\newcommand{\np}{\rlap{/}p}
\newcommand{\tr}{\mathop{\rm tr}}
\newcommand{\dd}[1]{\mathop{{\rm d}#1}}
\newcommand{\dds}[2]{\mathop{{\rm d\null}^{#1}#2}}

\let\epsilon\varepsilon
\let\bigl\Bigl
\let\bigr\Bigr

\newcommand{\nl}{\rlap{/}l}
\newcommand{\nb}{\rlap{/}b}
\renewcommand{\narrowtext}{\begin{multicols}{2} \global\columnwidth20.5pc}
\renewcommand{\widetext}{\end{multicols} \global\columnwidth42.5pc}
\def\top#1{\vskip #1\begin{picture}(290,80)(80,500)\thinlines \put(
65,500){\line( 1, 0){255}}\put(320,500){\line( 0, 1)
{ 5}}\end{picture}}
\def\bottom#1{\vskip #1\begin{picture}(290,80)(80,500)\thinlines \put(
330,500){\line( 1, 0){255}}\put(330,500){\line( 0, -1){
5}}\end{picture}}

\begin{document}

\title{Radiatively induced Lorentz and CPT violation in electrodynamics}

\author{R.\ Jackiw$^a$ and V.\ Alan Kosteleck\'y$^b$}

\address{$^a$Center for Theoretical Physics,
Massachusetts Institute of Technology, Cambridge, MA 02139-4307\\
$^{b}$Physics Department, Indiana University, Bloomington,
Indiana 47405}

\date{preprint IUHET 400, MIT CTP 2812, January 1999}

\maketitle

\begin{abstract}\noindent
In a nonperturbative formulation,
radiative corrections arising from Lorentz and CPT violation 
in the fermion sector induce a definite 
and nonzero Chern-Simons addition 
to the electromagnetic action.
If instead a perturbative formulation is used,
an infinite class of theories 
characterized by the value of the Chern-Simons coefficient
emerges at the quantum level.

\end{abstract}

\narrowtext
Lorentz and CPT symmetry of Maxwell electrodynamics is destroyed by
adding the term 
${\cal{L}}_k = \half k_\mu \e^{\mu\al\bb\g} F_{\al\bb} A_\g$
to the Lagrange density\cite{cfj,ck1,ck2}. 
Here,
$k_\mu$ is a prescribed, constant 4-vector. 
The term ${\cal{L}}_k$ is of the Chern-Simons form\cite{jt}: 
it changes by a total derivative 
when the gauge potential undergoes a gauge transformation 
$A_\mu \to A_\mu + \partial_\mu \Lambda$. 
Consequently,
the action and equations of motion are gauge invariant, 
but the Lagrange density is not. 
The modified theory predicts birefringence of light 
{\it in vacuo}\cite{cfj,ck2}.
Observation of distant galaxies 
puts a stringent bound on $k_\mu$:
it should effectively vanish\cite{cfj,r3}. 

A natural question is whether such a term would be induced through
radiative corrections when Lorentz and CPT symmetries are violated in
other sectors of a larger theory. 
If so,
then the stringent experimental limits on ${\cal{L}}_k$
would severely restrict the viability of models with 
Lorentz and CPT breaking\cite{fn1}. 

To study explicitly this issue, 
one may consider extending the quantum-electrodynamics (QED) action 
of a single Fermi field
by including a Lorentz- and CPT-violating axial-vector term\cite{ck1,ck2}:
\beq
I \supset \int \dds4x \bar\psi (i\npa -\nA  - m - \nb\g_5 )\psi\ .
\label{action}
\eeq
Here,
$b_\mu$ is  a constant, prescribed 4-vector, 
and our $\g_5$ is Hermitian with 
$\tr \g_5\g^\al \g^\bb \g^\g \g^\del = 4i \e^{\al\bb\g\del}$. 
The only other possible nonderivative CPT- and Lorentz-violating term
in the fermion sector is uninteresting here because 
its properties under charge conjugation prevent it contributing
to the Chern-Simons term\cite{ck2}.

Several calculations have been performed to determine whether
radiative corrections induce the Chern-Simons term with 
$k_\mu \propto b_\mu$.
At leading order in $b_\mu$ and the fine-structure constant,
a perturbative treatment of the term in Eq.~(\ref{action})
has been shown to generate an ambiguous result:
the coefficient $k_\mu$ of the induced Chern-Simons term
is regularization dependent and can be freely selected\cite{ck2}.
Other claims include both a 
definite zero value for $k_\mu$\cite{r5}
and a definite nonzero value\cite{co}.

The purpose of this work is to clarify this situation
and call attention to subtle issues,
related to chiral anomalies\cite{r6}, 
underlying the discrepancies between these various results.
First,
a direct calculation of $k_\mu$
is presented that is nonperturbative in $b_\mu$.
The various issues are then disentangled.

The relevant quantity for deciding whether a Chern-Simons term is
induced is the vacuum persistence amplitude,
or equivalently from Eq.~(\ref{action}) the fermion determinant
$\det (i\npa -\nA  - m - \nb\g_5 )$,
computed to second order in the photon variables. 
We are thus led to examining the standard one-loop
vacuum-polarization amplitude $\Pi^\mn$, 
but with the usual free-fermion propagator $S(l )$ 
replaced by the $b_\mu$-exact propagator from Eq.~(\ref{action}):
\beq
G(l) = \frac {i}{\nl - m -\nb \g_5} \ .
\label{eq:1}
\eeq
This may also be presented as 
\begin{mathletters}
\beq
G(l) = S(l) + G_b(l) \ , 
\label{eq:2a}
\eeq
where 
\beq
G_b(l) = \frac 1{\nl - m -\nb \g_5} \nb \g_5 S(l)\ .
\label{eq:2b}
\eeq
\end{mathletters}
With this decomposition,
$\Pi^\mn$ splits into three terms:
\beq
\Pi^\mn = \Pi^\mn_0 + \Pi^\mn_b + \Pi^\mn_{bb} \ .
\label{eq:3}
\eeq
The term $\Pi^\mn_0$
is the usual lowest-order vacuum-polarization tensor of QED, 
which we shall not discuss further. 
The term $\Pi^\mn_{bb}$ is at least quadratic in~$b$; 
it is at most logarithmically divergent 
and suffers no ambiguity in routing the internal momenta\cite{fn2}. 
The $b_\mu$-linear contribution to the Chern-Simons term 
arises from $\Pi^\mn_b$, 
which is given explicitly by 
\beqar
\Pi^\mn_b (p) &=& \int \frac{\dds4l}{(2\pi)^4} \tr \bigl\{
\g^\mu S(l) \g^\nu G_b (l +p) 
\nonumber \\ &&
\qquad\qquad\qquad
+ \g^\mu G_b(l ) \g^\nu S(l +p)\bigr\}\ .
\label{eq:4}
\eeqar

There are several important features of this expression.
Each of the two integrals is (superficially) linearly divergent.
However, 
the divergences cancel when the two terms are taken together 
and the traces are evaluated.
As a consequence, 
there is no momentum-routing ambiguity in the summed integrand: 
when the integration momentum is shifted by the same amount 
in {\it both} integrands the value of the integral does not change,
even though shifting separately by different amounts 
in each of the two integrands changes the value of the integral 
by a surface term. 
It follows that different momentum routings 
leave unchanged the value of $\Pi^\mn_b$ 
because they produce a simultaneous shift of integration variable 
by the same amount in each of the integrands in Eq.~(\ref{eq:4}). 
Therefore, 
a unique value can be calculated for $\Pi^\mn_b$, 
which we shall show leads to a finite Chern-Simons term.

We next evaluate $\Pi^\mn_b$ to lowest order in~$b$,
by replacing $G_b(l )$ with $-i S(l ) \nb \g_5 S(l )$. 
This gives 
\beq
\Pi^\mn_b \simeq \Pi^{\mu\nu\al} b_\al \ ,
\eeq
where
\beqar
 \Pi^{\mu\nu\al} (p) &=& -i \int\!\! \frac{\dds4l}{(2\pi)^4} \tr
\bigl\{
\g^\mu S(l ) \g^\nu S(l +p) \g^\al \g_5 S(l +p)
\nonumber\\ &&
\qquad\qquad\qquad\quad
+ \g^\mu S(l ) \g^\al \g_5 S(l ) \g^\nu S(l +p) \bigr\}
\nonumber\\
&\equiv& I^{\mn\al} (p) + \tilde I^{\mn\al}(p)\ .
\label{eq:6}
\eeqar

A shift of integration variables in the second term 
reduces it to a crossed form of the first
plus a contribution arising 
from shifting variables in a linearly divergent integral:
\beq
\tilde I^{\mn\al}(p) = I^{\nu\mu\al} (-p) + \Delta^{\mn\al} (p) \ ,
\eeq
where
\beqar
\Delta^{\mn\al} (p) &=& 
-i \int\!\! \frac{\dds4l}{(2\pi)^4} \tr \bigl\{
\g^\mu S(l ) \g^\al \g_5 S(l )\g^\nu S(l +p) 
\nonumber\\ &&
\qquad
- \g^\nu S(l ) \g^\mu S(l -p) \g^\al \g_5 S(l -p)\bigr\}\ .
\label{7}
\eeqar
Since the tensorial form of $I^{\mn\al}$ must be 
$\e^{\mn\al\bb} p_\bb$, 
the crossed term coincides with the uncrossed term. 
The variable shift produces a surface term, 
and $\Delta^{\mn\al}(p)$ is evaluated as 
\beq
\Delta^{\mn\al}(p) = -\frac1{8\pi^2} \e^{\mn\al\bb} p_\bb\ .
\label{eq:8}
\eeq
Thus,
we have  
\beqar
&&\Pi^{\mn\al} (p) =
-\frac1{8\pi^2} \e^{\mn\al\bb} p_\bb 
\nonumber\\ 
&& 
-2 \int\!\! \frac{\dds4l}{(2\pi)^4}
\tr \g^\mu \frac1{\nl-m} \g^\nu \frac1{\nl +\np -m}
\g^\al \g_5 \frac1{\nl +\np -m}.
\label{eq:9}
\eeqar
To evaluate the integral, 
note first that
\widetext
\top{-2.8cm}
\hglue -1 cm
\beqar
\frac1{\nl +\np -m} \g^\al \g_5 \frac1{\nl +\np -m} &=&
\frac1{\nl +\np -m}\g^a  \Bigl[ \frac{-1}{\nl +\np -m}
+ \frac{2m}{(l +p)^2-m^2} \Bigr]\g_5 
\nonumber\\ &=& 
\frac\pa{\pa p_\al} \frac1{\nl +\np -m}\g_5 + 
\frac{2m}{\nl +\np -m} \g^\al \g_5 \frac{1}{(l +p)^2-m^2} \ .
\label{eq:10}
\eeqar
The $p_\al$-derivative contributes a term 
$$
-2\frac\pa{\pa p_\al} \int \frac{\dds4l}{(2\pi)^4} \tr \Bigl(
\g^\mu \frac1{\nl-m} \g^\nu \frac1{\nl +\np -m} \g_5
\Bigr)\ . 
$$
However, 
the above integral must vanish:
no two-index pseudotensor exists 
involving the antisymmetric pseudotensor
and depending only on a single variable~$p$.
Therefore,
one is left with an entirely finite integral.
We find:
\beqar
\Pi^{\mn\al}(p) &=& 
-\frac1{8\pi^2} \e^{\mn\al\bb} p_\bb 
-4m \int \frac{\dds4l}{(2\pi)^4} \tr 
\g^\mu \frac1{\nl-m} \g^\nu \frac1{\nl +\np -m}
\g^\al \g_5 \frac1{(l +p)^2 - m^2} 
\nonumber\\
&=& 
-\frac1{8\pi^2} \e^{\mn\al\bb} p_\bb
+\frac{im^2}{\pi^4} \e^{\mn\al\bb}
p_\bb \int \dds4l \frac1{l ^2-m^2} 
\frac1{\left( (l +p)^2 - m^2\right)^2}
\nonumber\\
&=& -\e^{\mn\al\bb} p_\bb \Bigl(
\frac1{8\pi^2} + \frac2{\pi^2} \int_{2m}^\infty \dd a
\frac{m^2}{\sqrt{a^2-4m^2}}  \frac1{p^2-a^2+i\e}
\Bigr)
\label{eq:11a}
\eeqar
\bottom{-2.7cm}
\narrowtext
\noindent
The final result is:
\beqar
\Pi^{\mn\al}(p) &=& \e^{\mn\al\bb} \frac{p_\bb}{2\pi^2} \Bigl(
\frac\theta{\sin \theta} - \frac14 \Bigr) \ ,
\label{eq:11b}
\eeqar
where
$\theta \equiv 2\sin^{-1} (\sqrt{p^2}/2m)$
and $p^2 < 4m^2$.

The $b_\mu$-linear contribution to the induced Chern-Simons term 
is determined by this expression at $p^2=0$.
One finds a definite, nonzero, and finite result\cite{co,fn3}:
\beq
\left.\Pi^{\mn\al}(p)\right|_{p^2=0} = \frac3{8\pi^2} \e^{\mn\al\bb}
p_\bb \ ,
\label{eq:13}
\eeq
and 
\beq
k^\mu = \frac3{16\pi^2} b^\mu\ .
\label{eq:14}
\eeq
This completes our calculation.

We have chosen to extract the leading-order result in $b_\mu$.
However,
our calculation is in fact nonperturbative in $b_\mu$ 
in the sense that it has been performed
keeping careful track of contributions 
from the $b_\mu$-exact propagator in Eq.~(\ref{eq:2a}).
Thus, 
in this calculation,
we are choosing to define the theory 
of Eq.~(\ref{action}) in a nonperturbative way.

If instead the theory in Eq.~(\ref{action})
is defined through its perturbation series in $b_\mu$,
the same $\theta$ dependence as in Eq.~(\ref{eq:11b})
emerges but the additional constant and therefore the 
net result for the induced Chern-Simons term is different\cite{ck2}.
At first order,
one finds that the two integrals~(\ref{eq:6}) 
arise from a triangle VVA graph and its crossed expression
with zero axial-vector momentum. 
In perturbation theory, 
no correlation is determined {\it a~priori} 
between the momentum routings in the two graphs. 
If the relative routings are as in~(\ref{eq:6}), 
the resulting expression coincides with~(\ref{eq:11b}).
Otherwise,
a shift of integration variables produces 
the configuration~(\ref{eq:6}), 
but generates an additional contribution. 
Taking the shift as proportional to the external momentum
gives rise to an arbitrary multiple of 
$\Delta^{\mn\al} \propto \e^{\mn\al\bb} p_\bb$, 
leaving the Chern-Simons coefficient $k^\mu$
proportional to $b^\mu$
but with an {\it undetermined} proportionality constant. 

Coleman and Glashow have recently argued that 
$k_\mu$ must unambiguously vanish to first order in $b_\mu$
for any gauge-invariant CPT-odd interaction\cite{r5}.
Their result is based on their hypothesis 
that one define the axial vector 
$j^\mu_5 (x) \equiv \bar\psi{(x)}\g^\mu \g_5 \psi{(x)}$
to be gauge invariant in the quantum theory 
at arbitrary 4-momentum, that is, at every point in~$x$.
However, 
a weaker condition is true:
if $ j^\mu_5 (x)$ does not couple to any other field,
then physical gauge invariance is maintained 
provided $ j^\mu_5 (x)$ is gauge invariant at {\it zero} 4-momentum. 
Equivalently,
physical gauge invariance is maintained
provided $\int\dds4x j^\mu_5(x)$ and therefore the action
are gauge invariant, 
{\it without} the requirement that
the Lagrange density also be gauge invariant.
This behavior characterizes the Chern-Simons term, 
so it is unsurprising that demanding gauge invariance 
of the Lagrange density
can prevent generating the noninvariant Chern-Simons term. 

The Coleman-Glashow argument is perturbative in~$b_\mu$
and is taken to first order. 
Only in the perturbative framework does the axial vector 
arise as a distinct entity: 
it is an insertion whose gauge variance can be discussed. 
In contrast,
with the nonperturbative definition of the theory 
the axial vector has no separate identity, 
but when the first-order contribution 
is extracted from our complete expression
we find a nonzero result.
Evidently the dynamics of the nonperturbative theory 
selects the weaker option:
gauge invariance only for $\int\dds4x b_\mu j^\mu_5(x)$ 
but not for the unintegrated quantity.  
Gauge invariance is preserved in the sense that 
$p_\mu \Pi^\mn_b = 0$, 
and the induced action is gauge invariant, 
but the induced Chern-Simons Lagrange density is not. 

The gauge anomaly vanishes for zero momentum in the axial-vector vertex, 
regardless of the momentum routing in the two triangle graphs.
However,
for {\it nonvanishing} momentum in the axial vertex,
only a special routing of the integration momenta
gives a gauge-invariant answer.
This special routing is known explicitly\cite{r7},
and in the limit of zero axial-vector momentum 
it corresponds to the result of the Coleman-Glashow assumption,
giving uniquely
\beqar
\Pi^{\mn\al}_{\rm CG} (p) &=& -i \int
\frac{\dds4l}{(2\pi)^4} \tr \bigl\{
\g^\mu S(l ) \g^\nu S(l +p) \g^\al \g_5 S(l +p)\nonumber\\
 && \quad {}+
\g^\mu S(l +3p) \g^\al \g_5 S(l +3p) \g^\nu S(l +4p)
\bigr\}
\nonumber
\eeqar
after an inocuous shift in both integrands.
The integration momentum in the second expression 
must be decreased by $3p$ to bring 
this result into conformity with Eq.~(\ref{eq:6}). 
Therefore, 
from (\ref{eq:8}) it follows that
\beqar
\Pi^{\mn\al}_{\rm CG} (p) &=&
\Pi^{\mn\al} (p) - \frac3{8\pi^2}
\e^{\mn\al\bb} p_\bb
\ .
\label{eq:15b}
\eeqar
This result vanishes at $p^2=0$,
in agreement with the Coleman-Glashow claim. 
However,
other than the demand that gauge invariance be maintained 
not only for the action but also for the induced Lagrange density, 
there is no reason to prefer this result over any other. 

A further degree of arbitrariness in the induced term
appears according to the choice of regularization scheme
used in the calculation.
This can be true even within a nonperturbative formulation.

One possible choice is Pauli-Villars regularization,
which enforces gauge invariance for all axial momenta. 
The various values of the induced Chern-Simons coefficient 
are mass-independent,
so they are subtracted and vanish 
in Pauli-Villars regularization\cite{ck2}. 
This would be true whether or not
the theory is formulated perturbatively or nonperturbatively. 
Another possibility is dimensional regularization.
This is problematic with the $\g_5$-matrix, 
and a variety of answers for $k_\mu$ can be obtained
depending on how $\g_5$ is generalized to arbitrary dimensions.
In this sense,
the physics of the theory (\ref{action})
depends on the choice of regularization scheme.

Referring to the dispersive representation,
presented in Eq.~(\ref{eq:11b}), 
we see that the theory predicts a definite absorptive part, 
which is sufficiently well behaved to enter 
into an unsubtracted dispersion relation. 
Nevertheless, 
there remains the possibility of a real subtraction, 
which is perturbatively undetermined, 
even though we presented a nonperturbative
argument for the value $1/8\pi^2$. 
This situation is familiar in quantum field theory.
Parameters in a Lagrangian typically
acquire infinite radiative corrections that must be renormalized. 
They flow with the renormalization scale and are determined
only by comparison with experiment.  
Here, 
the radiative corrections are finite,
so infinite renormalization is unnecessary,
but nevertheless no definite value is determined
in perturbation theory. 
Another instance of this phenomenon occurs 
in the chiral Schwinger model, 
which generates an undetermined mass 
for its vector meson\cite{r8}. 
The $\theta$-angle of QCD provides a further example\cite{r9}. 
For all these, 
a finite parameter must be fixed by reference to Nature.
For the Chern-Simons case this has already been done\cite{cfj,r3}. 

In this work,
we have found that the apparently reasonable physical question 
``Is a Chern-Simons term induced in the theory (\ref{action})?''
has no unique answer.
We have given a nonperturbative definition 
of the theory that induces a definite and nonzero value 
of the Chern-Simons coefficient. 
If instead a perturbative definition is used,
an infinite class of theories 
characterized by the value of the Chern-Simons coefficient
emerges at the quantum level.
The choice of regularization procedure can induce further ambiguity
in both nonperturbative and perturbative schemes.
Although one could perhaps argue that 
our nonperturbative formulation 
is the most aesthetically satisfying,
there seems no compelling reason to prefer
any one definition over another.

\bigskip

We thank Don Colladay, J.-M.\ Chung, and P.\ Oh
for useful discussions.
This work is supported in part by 
the United States Department of Energy under grant numbers
DE-FC02-94ER40818 and DE-FG02-91ER40661.

\end{multicols}

\begin{thebibliography}{99}
\bibitem{fn1}
Even if a nonzero term is induced by a particular sector
of a larger theory,
the net effect can still vanish.
This would occur in the standard-model extension
of Ref.~\cite{ck2}
through the anomaly-cancellation mechanism,
for example.
\bibitem{fn2}
In fact,
the contribution from $\Pi^\mn_{bb}$ 
to the induced Chern-Simons term
is at least cubic in~$b$ and hence finite
because all terms quadratic in~$b$ are necessarily CPT even.
\end{thebibliography}
\end{document}